\documentclass[sigconf,nonacm]{acmart}
\AtBeginDocument{%
  \providecommand\BibTeX{{%
    \normalfont B\kern-0.5em{\scshape i\kern-0.25em b}\kern-0.8em\TeX}}}

\setcopyright{acmcopyright}
\copyrightyear{2022}
\acmYear{2022}
\acmDOI{XXXXXXX.XXXXXXX}

\acmConference[SCORES'25]{Student Computing Research Symposium}{September 10,
2025}{Ljubljana, Slovenia}
\acmPrice{15.00}
\acmISBN{978-1-4503-XXXX-X/18/06}

\usepackage{verbatim}
\usepackage[linesnumbered, boxed, resetcount]{algorithm2e}
\usepackage{tikz}
\usetikzlibrary{calc}
\usetikzlibrary{backgrounds}

\begin{document}

\title{Empirical Analysis Of Heuristic And Approximation Algorithms For The Mutual-Visibility Problem}

\author{Vanja Stojanovi\'c}
\authornote{Both authors contributed equally to this research.}
\email{vs66277@student.uni-lj.si}
\affiliation{%
    \institution{Faculty of Mathematics and Physics,\\ University of Ljubljana\\ Jadranska ulica 21}
    \city{SI-1000 Ljubljana}
    \country{Slovenia}
}

\author{Bor Panger\v si\v c}
\email{bp52226@student.uni-lj.si}
\affiliation{%
    \institution{Faculty of Mathematics and Physics,\\ University of Ljubljana\\ Jadranska ulica 21}
    \city{SI-1000 Ljubljana}
    \country{Slovenia}
}

\begin{abstract}
The NP-complete mutual-visibility (MV) problem currently lacks empirical analysis on its practical behaviour despite theoretical studies. This paper addresses this gap by implementing and evaluating three distinct algorithms -- a direct random heuristic, a hypergraph-based approximation, and a genetic algorithm -- on diverse synthetic graph datasets, including those with analytically known $\mu(G)$ values and general graph models. Our results demonstrate that for smaller graphs, the algorithms consistently achieve MV set sizes aligning with theoretical bounds. However, for larger instances, achieved solution sizes notably diverge from theoretical limits; this, combined with the absence of tight bounds, complicates absolute quality assessment. Nevertheless, validation on known optimal graphs showed the Genetic Algorithm and other heuristics empirically performing best among tested methods.
\end{abstract}


\maketitle

\section{Introduction}

All code and datasets for this work are available at our
\href{https://github.com/Vanja-S/SCORES-25---Construction-Of-A-Greedy-Algorithm-And-Its-Empirical-Analysis-On-The-MV-Problem}{GitHub repository}.

The mutual-visibility (MV) problem was introduced by Di Stefano et al.\ in \cite{DISTEFANO2022126850} and was subsequently generalized and extended in \cite{CICERONE2023114096}. Given a subset \( X \subseteq V(G) \), we say that vertices \( u, v \in X \) are \emph{X-visible} if there exists a shortest path \( P \) such that \(V(P) \cap X \subseteq \{u, v\} \). Such a set \( X \) is called a \emph{mutual-visibility set} if every pair \( u, v \in X \) are X-visible. A $\mu$-set is a maximum mutual-visibility set (i.e., no larger mutual-visibility set exists). When we only require that no proper superset is an mutual-visibility set, we speak of a maximal MV-set. Let $\mu(G)$ denote the cardinality of a $\mu$-set of graph G.

The mutual-visibility decision problem is stated as follows; given a graph \(G = (V,E)\) and set \(P \subseteq V(G)\), is \(P\) a mutual-visibility set? Di Stefano derived and proved a polynomial-time BFS-based algorithm for solving this problem. However, in general, finding the largest such set, i.e. finding the \emph{mutual-visibility number} of a graph, is proved to be NP-complete \cite{DISTEFANO2022126850}.

For the mutual-visibility and related problems, Bilò et al. proved strong inapproximability results, demonstrating that it is not approximable within $n^{1/3-\epsilon}$ for graphs of diameter at least 3, and is APX-Hard for graphs of diameter 2. While exact solutions for some specific graph classes are known, there is also a polynomial-time approximation algorithm for general graphs \cite{bilò2024approximabilitygraphvisibilityproblems}. Its precise lower bound on the size of the found set is discussed in detail in the \hyperref[sec:approximation_algo_using_hypergraphs]{Approximation Algorithm Using Hypergraphs} section.
\subsection{Motivation}
Despite the theoretical analysis of the MV problem, there have been no public attempts to empirically analyse its behaviour in practice or to systematically implement and evaluate various heuristic and approximation algorithms. This paper addresses this gap by conducting a comprehensive empirical analysis of the MV problem. Our primary motivations are threefold:
\begin{enumerate}
    \item To gain practical insights into the difficulty of approximating the mutual-visibility number, moving beyond theoretical bounds to observe real-world performance.
    \item To empirically verify if established theoretical upper and lower bounds are observed in practice.
    \item To comparatively evaluate the performance of different algorithmic approaches.
\end{enumerate}
To achieve these objectives, we develop and analyse three distinct algorithms: a direct random heuristic, an approximation algorithm based on a hypergraph transformation, and a metaheuristic approach using a Genetic Algorithm. Our empirical evaluation is conducted on a dataset that includes graphs with analytically known mutual-visibility numbers (for validation) as well as other synthetic graph models.

\section{Algorithms and Heuristics}

\subsection{Direct Random Algorithm}
We construct a simple random algorithm for finding the largest mutual-visibility set within a graph $G$. The algorithm works by random sampling; by first choosing a subset size $k$ uniformly from $1..n$, then sampling a $k$-subset of $V(G)$, and checking for the mutual-visibility property of the set. As a default the number of trials $T$ was set to $10,000$. The algorithm is described in Algorithm \ref{algo:random_mv_sampling_size}. Note that the procedure $\text{MV}(G, S_{candidate})$ is defined in \cite[p.~7]{DISTEFANO2022126850} as Procedure MV. 

\DontPrintSemicolon
\begin{algorithm}
    \small
    \DontPrintSemicolon
    \SetAlgoLined
    \KwIn{Graph $G$, number of trials $T$}
    \KwOut{Best mutual‐visibility set $S^*$}
    $n \gets |V(G)|$\;
    $S^* \gets \emptyset$\;
    \For{$i \gets 1$ \KwTo $T$}{
        $k \gets \text{random integer in }[1,n]$\;
        $S \gets \text{uniformly random sample of $k$ nodes from }V(G)$\;
        \If{$|S| > |S^*|\ \land\ \mathrm{MV}(G,S)$}{
            $S^* \gets S$\;
        }
    }
    \Return{$S^*$}\;
    \caption{Random Sampling by Subset Size for Mutual‐Visibility}
    \label{algo:random_mv_sampling_size}
\end{algorithm}

\subsection{Genetic Algorithm (GA)}
We adapt a GA approach to find a large mutual-visibility set, drawing inspiration from its application to the General Position Problem in \cite{hamedlabbafian2025algorithmicapproachesgeneralposition}.

Each candidate MV set is represented as an \textbf{individual} in the form of a binary characteristic vector of size $n$, where a `1` at index $i$ indicates the inclusion of node $i$, and `0` indicates exclusion. The algorithm's \textbf{fitness function} evaluates each individual $S$ using the formula: $\text{Fitness}(S) = |S| - M \cdot f$. Here, $|S|$ is the number of selected vertices, $M$ is a large penalty constant, and $f$ represents the number of violations. A violation occurs if a pair of vertices within $S$ are not mutually visible (i.e., there exists no shortest path between them free of other vertices in $S$).

The algorithm begins with an \textbf{initial population} of individuals generated randomly (the solutions are not necessarily feasible). In each generation, \textbf{parent solutions} are randomly selected, and a \textbf{crossover operator} is applied (a bitwise OR operation). A \textbf{mutation operator} works by randomly selecting two gene positions (vertices) in an individual's characteristic vector and swapping their values. The \textbf{population update} phase merges the current population with the newly generated offspring and mutated individuals, sorts them by fitness, and truncates the population to maintain the predefined size, forming the next generation. The \textbf{stopping criterion} is defined as a maximum number of iterations.

The general structure of the adapted GA is outlined in Algorithm \ref{algo:ga_mv}. The parameters $p_c$ and $p_m$ were both chosen simply as $1.0$.

\begin{algorithm}[ht]
\small
\SetAlgoLined
\KwIn{Graph $G$; maximum iterations $T$; population size $N$; crossover probability $p_c$; mutation probability $p_m$; penalty parameter $M$}
\KwOut{Best mutual-visibility set $S^*$ found}
Initialize a population $\mathcal{P}$ of $N$ individuals\;
Evaluate the fitness of each individual in $\mathcal{P}$\;
\For{$t \gets 1$ \KwTo $T$}{
    Select parents from $\mathcal{P}$\;
    Generate offspring via crossover with probability $p_c$\;
    Apply mutation to offspring with probability $p_m$\;
    Form a new population by merging parents, offspring, and mutants\;
    Evaluate fitness of all individuals in the new population\;
    Select the best $N$ individuals to form the next generation $\mathcal{P}$\;
}
$S^* \gets$ individual in $\mathcal{P}$ with the highest fitness\;
\Return{$S^*$}\;
\caption{Genetic algorithm for approximating a mutual-visibility set}
\label{algo:ga_mv}
\end{algorithm}

\subsection{Approximation Algorithm Using Hypergraphs}
\label{sec:approximation_algo_using_hypergraphs}
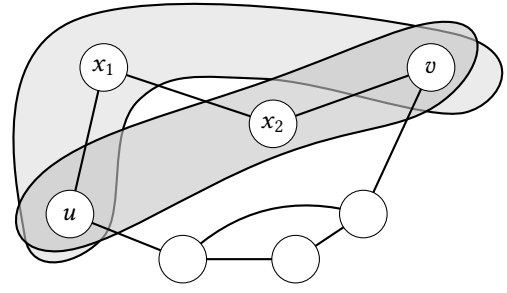
\begin{figure}
    \begin{tikzpicture}[scale=1.5, every node/.style={circle,draw,fill=white,minimum size=18pt,inner sep=1pt, font=\large}]
    
    \coordinate (u) at (0,0);
    \coordinate (x1) at (0.3,1.3);
    \coordinate (x2) at (1.8,0.8);
    \coordinate (v) at (3.2,1.3);
    \coordinate (a) at (1,-0.4);
    \coordinate (b) at (2,-0.4);
    \coordinate (c) at (2.6,0);
    
    \begin{pgfonlayer}{background}
        \filldraw[gray!25, fill opacity=0.6,, draw=black, thick] plot[smooth cycle, tension=0.7] coordinates {
            ($(u)+(-0.4,-0.2)$) ($(x1)+(-0.3,0.4)$)
            ($(v)+(0.2,0.3)$) ($(v)+(0.3,-0.4)$)
            ($(x2)+(-0.1,0.4)$) ($(x1)+(0.3,-0.3)$)
            ($(u)+(0.3,-0.2)$)
        };
        
        \fill[gray!45, fill opacity=0.6,, draw=black, thick] plot[smooth cycle, tension=0.8] coordinates {
            ($(u)+(-0.2,0.4)$) ($(x2)+(0,0.4)$)  
            ($(v)+(0.2,0.4)$) ($(v)+(0.1,-0.3)$) ($(x2)+(0,-0.3)$)
            ($(u)+(-0.1,-0.33)$)
        };
        
    \end{pgfonlayer}
    
    \draw[thick] (u) -- (x1);
    \draw[thick] (x1) -- (x2);
    \draw[thick] (x2) -- (v);
    \draw[thick] (u) -- (a);
    \draw[thick] (a) to[bend left=30] (c);
    \draw[thick] (a) -- (b);
    \draw[thick] (b) -- (c);
    \draw[thick] (c) -- (v);
    
    \node at (u) {$u$};
    \node at (x1) {$x_1$};
    \node at (x2) {$x_2$};
    \node at (v) {$v$};
    \node[draw,fill=white] at (a) {};
    \node[draw,fill=white] at (b) {};
    \node[draw,fill=white] at (c) {};
    
    \end{tikzpicture}
    \caption{Generation of the hypergraph}
    \label{fig:hypergraph}
\end{figure}

The approximation algorithm leverages a transformation of the MV problem into finding an independent set in a 3-uniform hypergraph, as described in \cite{bilò2024approximabilitygraphvisibilityproblems}. 
The hypergraph $H$ is constructed with the same set of vertices as the original graph $G$, i.e., $V(H) = V(G)$. For any unordered pair of distinct vertices $\{u, v\} \subseteq V(G)$, a shortest path $\langle u, x_1, x_2, \ldots, x_k, v \rangle$ is considered. The set of internal vertices on this path is denoted $B(\{u, v\}) = \{x_1, \ldots, x_k\}$, which can be empty if $k=0$ (i.e., $u$ and $v$ are adjacent). A hyperedge $\{u, v, x\}$ is then added to $H$ for each such pair $\{u, v\}$ and for every internal vertex $x \in B(\{u, v\})$. See Figure \ref{fig:hypergraph} for an example.

An independent set $S$ of this hypergraph $H$ is a subset of $V(G)$ such that no hyperedge $e$ (e.g., $\{u, v, x\}$) is entirely contained within $S$. Notice, an independent set $S$ of $H$ is exactly a $\mu$-set of $G$. That is, if $u, v \in S$ are two distinct vertices, it must be the case that no vertex $x \in B(\{u, v\})$ is also in $S$, as $H$ would otherwise contain the hyperedge $\{u, x, v\}$. The methodology assumes that for graphs with at most 6 vertices, a $\mu$-set can be computed by brute force.

Once the hypergraph $H$ is constructed, the algorithm proceeds to compute an independent set $S$ of $H$ using a greedy selection strategy defined in \cite{CARO1991}. The general approach for this phase is outlined in Algorithm \ref{algo:hypergraph_greedy}.

\begin{algorithm}[ht]
\small
\SetAlgoLined
\KwIn{Graph $G$}
\KwOut{Approximated mutual-visibility set $S$}
\tcp{\textbf{Phase 1: Hypergraph Construction}}
Construct a hypergraph $\mathcal{H}$ from $G$ as described in the text\;
\tcp{\textbf{Phase 2: Greedy Independent Set Selection}}
$S \gets V(H)$\\ 
$H' \gets H$\\ 
\While{There exists a vertex $y \in V(H')$ such that its degree $deg_{H'}(y) \ge 1$}{
    Select $y^* \in V(H')$ s. t. $deg_{H'}(y^*)$ is maximum\;
    $S \gets S \setminus \{y^*\}$\;
    Remove $y^*$ from $V(H')$\;
    Remove all edges incident to $y^*$ from $E(H')$\;
}
\Return{$S$}\;
\caption{Greedy algorithm for approximating an MV set via hypergraph independent set}
\label{algo:hypergraph_greedy}
\end{algorithm}
\normalsize

The approximation algorithm is particularly useful, because it is guaranteed to find a mutual-visibility set of size \(\Omega\Bigl (\sqrt{n / \overline {D}}\Bigr)\), where \(\overline D\) is the average distance between vertices of a graph, rigorously defined in \eqref{eq:avg_dist}. The proof of the bound lower is provided in \cite{bilò2024approximabilitygraphvisibilityproblems}.
\begin{equation}
    \label{eq:avg_dist}
    \overline D = \frac{2}{n(n-1)} \sum_{\{u,v\} \in \binom{V(G)}{2}} d(u,v)
\end{equation}

\begin{table*}[h!]
    \centering
    \footnotesize 
    \caption{Graph Classes for Empirical Evaluation.}
    \label{tab:dataset_graphs}
    \setlength{\tabcolsep}{4pt} 
    \hspace*{-0.3cm}
    \begin{tabular}{lp{5.8cm}ll} 
        \toprule
        \textbf{Graph Class} & \textbf{Key Properties / Purpose} & \textbf{Parameters Varied} & \textbf{Expected $\mu(G)$ (for validation)} 
        \cite{DISTEFANO2022126850},\cite{roy2024mutualvisibilitygeneralpositiondouble}\\
        \midrule
        \multicolumn{4}{l}{\textbf{Category 1: Analytical Solutions}} \\
        \multicolumn{4}{l}{\textit{These classes have simple, direct formulas for $\mu(G)$, or known tighter bounds obtained in recent research.}} \\
        Complete Graphs ($K_n$) & Densest, fully connected. & $n$ & $|V|$\\
        Trees & Sparse, acyclic, general case. & $n$ & $|L|$ (set of leaves)\\
        Grid Graphs ($\Gamma_{m,n}$) & Regular 2D lattice. & $m, n$ & $2 \cdot \min(m,n)$ for $m,n > 3$\\
        Tori Graphs ($T_{m,n}$) & Regular 2D lattice with wrap-around. & $m, n$ & $\le 3 \cdot \min(m,n)$ (exact for $m=12,15$) \\
        Mycielskian of Cycles ($M(C_n)$) & Mycielskian operation on cycle graphs. & $n$ (base) & $n + \lfloor\frac{n}{4}\rfloor$ for $n \ge 8$ (and $n+2$ for $4 \le n \le 7$)\\
        Mycielskian of Paths ($M(P_n)$) & Mycielskian operation on path graphs. & $n$ (base) & $n + \lfloor\frac{n+1}{4}\rfloor$ for $n \ge 5$ (and $6$ for $n=4$)\\
        Mycielskian of Star Graphs ($M(K_{1,k})$) & Mycielskian on stars (base has universal vertex). & $k+1$ (base nodes) & $2k+1$ for $k \ge 1$ (from $2n-1$ for base $n$)\\
        \midrule
        \multicolumn{4}{l}{\textbf{Category 2: General Graph Models}} \\
        Generalized Petersen Graphs ($G(n,k)$) & Specific highly symmetric cubic graphs. & $n, k$ & N/A (generally unknown/no simple formula for $\mu(G)$) \\
        Erd\H{o}s-R\'enyi ($G(n,p)$) & Varied density, random edge distribution. & $n, p$ & N/A (generally unknown/probabilistic) \\
        \bottomrule
    \end{tabular}
\end{table*}

\section{Experimental Methodology}

The algorithms will be executed on graph instances from the `n10` and `n100` size categories within the curated dataset. For each graph instance, both the Direct Random Heuristic, the Approximation Algorithm Using Hypergraphs, and the Genetic Algorithm will be run multiple times, with average results recorded.

\subsection{Data Sets}
The empirical analysis is conducted on a single, dataset of synthetically generated graphs. It comprises various graph classes, each selected to serve specific purposes in our analysis. Table \ref{tab:dataset_graphs} provides a detailed overview of these classes, including their key properties, the parameters varied during their generation, and their expected $\mu(G)$ values for validation purposes.
The graph classes are broadly categorized as follows:
\begin{enumerate}
    \item \textbf{Category 1: Analytical Solutions.} This category includes  graphs where $\mu(G)$ can be calculated directly using well-known formulas. It also features more complex graphs for which tighter bounds were obtained in recent research.
    \item \textbf{Category 2: General Graph Models.} This category includes less predictable graph structures, namely Generalized Petersen Graphs and connected Erd\H{o}s-R\'enyi graphs. For these models, $\mu(G)$ is generally unknown or computationally intractable to determine exactly.
\end{enumerate}

The reason for choosing these general graph categories is due to the simplicity of generating Erd\H{o}s-R\'enyi graphs, and the still analytically unknown solutions to the Generalized Petersen graphs. We hope to provide empirical insight into the structure of the MV problem on such variety of graphs.

\subsection{Performance Metrics and Analysis}
The primary metrics collected for each algorithm on every graph instance will include approximated MV set size $|S|$ as well as execution time.

For graph classes in Category 1 (where $\mu(G)$ is known analytically), the \textbf{approximation ratio} ($|S|/\mu(G)$) will be calculated.

The analysis will involve:
\begin{itemize}
    \item \textbf{Comparative Performance Across Scales:} Evaluating how the average approximation ratio and runtime of each algorithm vary across different size categories (`n10`, `n100`) and graph types (e.g., Complete, Tree, Grid, Mycielskian types, Erd\H{o}s-R\'enyi).
    \item \textbf{Bound Verification:} For graph classes with known theoretical bounds, plotting the achieved $|S|$ against the theoretical lower bounds (e.g., $\sqrt{n/\overline{D}}$, $\omega(G)$) to visually assess if these bounds are met or exceeded in practice.
\end{itemize}

\section{Results and Analysis}

\subsection{Overall Performance Overview}
We first provide a high-level summary of the algorithm's average performance for Category 1 of the dataset. 
Notice that in larger graph instances, the average performance of GA can occasionally suffer due to the algorithm getting stuck in local maxima. This results in a few individual cases exhibiting very poor approximation ratios, while other individual runs on similar instances might still perform well. These findings are visible in Tables \ref{tab:avg_ratio_summary}, where 
$\overline \alpha$ represents \textit{average ratio for the approach}, and \ref{tab:avg_performance_summary}, where $\overline T$ represents \textit{average runtime} for the approach. Notice that the genetic algorithm has $\overline{\alpha}$ over 1 in some cases, this is due to its indeterministic nature, where it can choose a bigger $S$ than $\mu(G)$, and it does not optimize it in the specified maximum iterations.

\begin{table}[H]
    \centering
    \footnotesize
    \caption{Average Ratio by Algorithm, Graph Type, and Size Category.}
    \label{tab:avg_ratio_summary}
    \setlength{\tabcolsep}{4pt} 
    \begin{tabular}{llcccc}
        \toprule
        \textbf{Graph Type} & \textbf{Size Category} & \textbf{$\overline \alpha$ Random} & \textbf{$\overline \alpha$ Hyper} & \textbf{$\overline \alpha$ Genetic} \\
        \midrule
        Complete & n10 & 1.000 & 1.000 & 1.000 \\
        Complete & n100 & 1.000 & 1.000 & 1.000 \\
        Grids & n10 & 0.892 & 0.675 & 0.767 \\
        Grids & n100 & 0.822 & 0.707 & 3.897 \\
        Mycielskian & n10 & 0.917 & 0.889 & 0.919 \\
        Mycielskian & n100 & 0.587 & 0.867 & 0.815 \\
        Trees & n10 & 1.000 & 1.000 & 1.024  \\
        Trees & n100 & 0.547 & 1.000 & 2.573 \\
        \bottomrule
    \end{tabular}
\end{table}

\vspace*{-0.4cm}

\begin{table}[H]
    \centering
    \footnotesize
    \caption{Average Performance by Algorithm, Graph Type, and Size Category.}
    \label{tab:avg_performance_summary}
    \setlength{\tabcolsep}{4pt} 
    \begin{tabular}{llcccc}
        \toprule
        \textbf{Graph Type} & \textbf{Size Category} & \textbf{$\overline T$ Random [s]} & \textbf{$\overline T$ Hyper [s]} & \textbf{$\overline T$ Genetic [s]} \\
        \midrule
        Complete & n10 & 0.014 & 0.000 & 0.015 \\
        Complete & n100 & 0.709 & 0.024 & 7.000 \\
        Grids & n10 & 0.280 & 0.001 & 0.228 \\
        Grids & n100 & 1.998 & 3.206 & 10.503 \\
        Mycielskian & n10 & 0.196 & 0.001 & 0.213 \\
        Mycielskian & n100 & 4.758 & 1.085 & 29.246 \\
        Trees & n10 & 0.129 & 0.000 & 0.025 \\
        Trees & n100 & 1.471 & 19.924 & 7.170 \\
        \bottomrule
    \end{tabular}
\end{table}

\subsection{Approximation Quality Analysis}
\begin{figure}[ht!]
    \centering
    \includegraphics[width=0.9\linewidth]{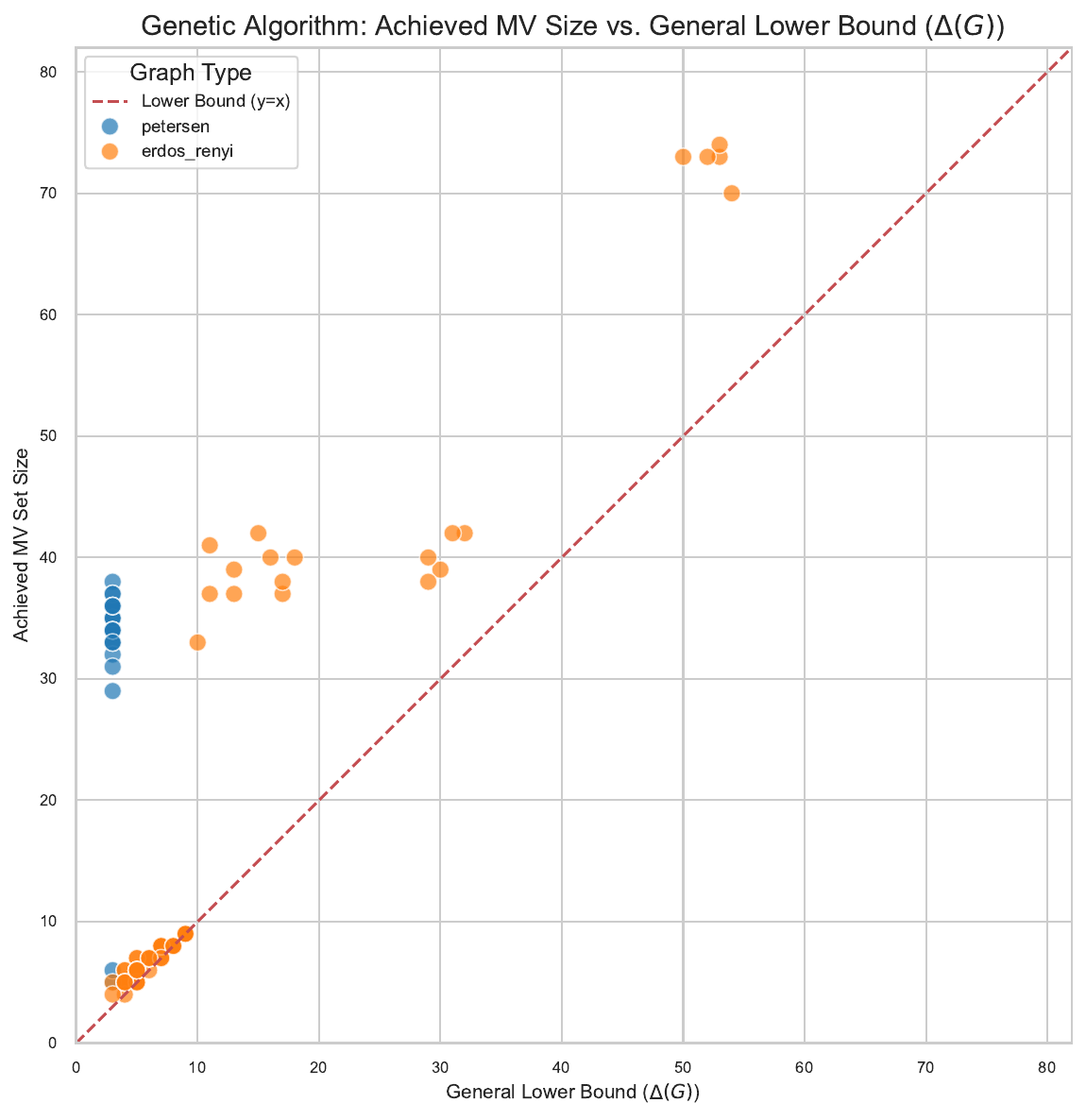}
    \caption{$\mu$-set size vs. general lower bound for genetic algorithm approach.}
    \label{fig:genetic_mv_vs_upper}
\end{figure}

Our empirical analysis reveals that for graphs with a small number of vertices (e.g., $n=10$), the obtained mutual-visibility set sizes consistently approximate or closely align with the established theoretical lower and upper bounds. However, for larger graph instances (eg. $n = 100$), this relationship becomes less evident. The random‐sampling algorithm provides a natural baseline against which to compare more complex methods for the mutual-visibility problem.  Empirical results show that, with \(T = 10\,000\) trials, it often yields near-optimal estimates of the mutual-visibility number \(\mu(G)\) on small (e.g.\ \(n=10\)) and even on certain larger graphs of certain classes.  However, as the order of the graph increases, the quality of the approximation deteriorates unless the number of random trials is increased proportionally. Another observation is that the lower bound obtained from the hypergraph approximation method is on average lower than the general lower bound provided by the largest degree and hence not very useful in providing tighter bounds. For the genetic algorithm approach the relation between the achieved $\mu$-set size vs. general lower bound ($\Delta(G)$) can be observed from the graph in Figure \ref{fig:genetic_mv_vs_upper}. Note that the blue coloured data points represent analysis on general Petersen graphs, while orange data points represent Erd\H{o}s-R\'enyi graphs.

\vspace*{-0.55cm}

\section{Conclusion}
This paper presented an empirical analysis of three distinct algorithmic approaches-- a direct random heuristic, a hypergraph-based approximation, and a Genetic Algorithm-- for solving the NP-complete MV problem. Our evaluation across various graph datasets revealed that for larger graph instances, the achieved solution sizes notably diverge from theoretical limits which, in the absence of tight bounds or known optimal values for larger graphs, definitively assessing the absolute quality of the found solutions becomes challenging. Nevertheless, from the validation performed on graph classes with analytically known $\mu(G)$, the Genetic Algorithm, and potentially other general heuristic approaches, empirically demonstrated superior performance among the tested methods.

\section{Future Work}
From the initial insights into the practical hardness of the MV problem, several promising directions for future research emerge.

Other than including other graph instances and a broader range of sizes to evaluate, a promising direction involves a deeper investigation into the correlation between specific graph properties and algorithmic performance. Beyond the general graph types studied here, future work could examine how various graph properties such as density, diameter, and average degree influence the achieved MV set size and runtime across different graph classes.

Furthermore, while this paper focused on a direct random heuristic, a hypergraph-based approximation, and a Genetic Algorithm, future research could explore other metaheuristics like Simulated Annealing, or exact methods for smaller instances, such as Integer Linear Programming, as demonstrated by authors in \cite{hamedlabbafian2025algorithmicapproachesgeneralposition}.

Finally, moving beyond theoretical graphs, a compelling avenue involves applying these algorithms to more real-world examples. This directly links back to the problem's initial motivation in robotic navigation, where algorithms for repositioning robots on a graph while maintaining mutual visibility are crucial \cite{DISTEFANO2022126850}.

\begin{acks}
    We thank Assoc. Prof. PhD Csilla Bujtás for her guidance during the initial stages of this research.
    
    We also extend our gratitude to Assist. Prof. Dr. Uroš Čibej for his valuable advice and support throughout the preparation of this paper. 
\end{acks}
\bibliographystyle{ACM-Reference-Format}
\bibliography{references}

\end{document}